\documentclass[conference, a4paper]{IEEEtran}
\IEEEoverridecommandlockouts
\usepackage[noadjust]{cite}
\usepackage{amsmath,amssymb,amsfonts}
\usepackage{bm}
\usepackage{algorithmic}
\usepackage[inline]{enumitem}
\usepackage{graphicx}
\usepackage{textcomp}
\usepackage{xcolor}
\usepackage{psfrag}

\usepackage{scalerel}
\usepackage{tikz}
\usetikzlibrary{svg.path}
\definecolor{orcidlogocol}{HTML}{A6CE39}
\tikzset{
  orcidlogo/.pic={
    \fill[orcidlogocol] svg{M256,128c0,70.7-57.3,128-128,128C57.3,256,0,198.7,0,128C0,57.3,57.3,0,128,0C198.7,0,256,57.3,256,128z};
    \fill[white] svg{M86.3,186.2H70.9V79.1h15.4v48.4V186.2z}
                 svg{M108.9,79.1h41.6c39.6,0,57,28.3,57,53.6c0,27.5-21.5,53.6-56.8,53.6h-41.8V79.1z M124.3,172.4h24.5c34.9,0,42.9-26.5,42.9-39.7c0-21.5-13.7-39.7-43.7-39.7h-23.7V172.4z}
                 svg{M88.7,56.8c0,5.5-4.5,10.1-10.1,10.1c-5.6,0-10.1-4.6-10.1-10.1c0-5.6,4.5-10.1,10.1-10.1C84.2,46.7,88.7,51.3,88.7,56.8z};
  }
}
\newcommand\orcidicon[1]{\href{https://orcid.org/#1}{\mbox{\scalerel*{
\begin{tikzpicture}[yscale=-1,transform shape]
\pic{orcidlogo};
\end{tikzpicture}
}{|}}}}

\usepackage{pgfplots}
\pgfplotsset{compat=newest}

\usepackage{hyperref} 
\usepackage{cleveref}

\crefformat{footnote}{#2\footnotemark[#1]#3}

\DeclareMathOperator{\diag}{diag}
\DeclareMathOperator{\spanS}{span}

\def\Expct{\mathbb{E}}
\def\Cmplx{\mathbb{C}}
\def\Real{\mathbb{R}}
\def\CNormal{\mathcal{CN}}
\def\vectorize{\mathrm{vec}}
\def\by{\bm y}

\def\bxt{\bm x_{\rm T}}
\def\bxc{\bm x_{\rm C}}
\def\bx{\bm x}
\def\bc{\bm c}
\def\bs{\bm s}
\def\bf{\bm f}
\def\bg{\bm g}

\def\bphi{\bm \phi}
\def\bsigma{\bm \sigma}
\def\be{\bm e}
\def\bk{\bm k}
\def\bA{\bm A}

\def\bF{\bm F}
\def\bS{\bm S}
\def\bAxt{\bA_{\rm T}}
\def\bAxc{\bA_{\rm C}}
\def\bSigma{\bm\Sigma}
\def\bSigmac{\bm\Sigma_{\rm C}}
\def\bP{\bm P}
\def\BFR{\mbox{BFR}}

\def\arXivPrint{1}  

\makeatletter
\newcommand\gobblestar
    {\futurelet\@let@token\@dogobblestar}

\def\@dogobblestar
    {\let\next=\relax
     \ifx\@let@token*%
        \def\next{\@gobble}%
     \else\ifx\@let@token[%
         \def\next{\@gobbleoptional}%
     \fi\fi
     \next}

\def\@gobbleoptional[#1]{}
\makeatother

\newcommand\doSingleLine[1]
    {\let\normalnewline=\\
     \let\\=\gobblestar
     #1%
     \let\\=\normalnewline}
\begin{document}

\title{Kernel Design Meets Clutter Cancellation\\for Irregular Waveforms}

\author{\IEEEauthorblockN{Pepijn B. Cox \orcidicon{0000-0002-8220-7050}, Mario A. Coutino \orcidicon{0000-0003-2228-5388}, and Wim L. van Rossum}
\IEEEauthorblockA{\textit{Radar Technology Department, TNO} \\
The Hague, The Netherlands \\
\{pepijn.cox, mario.coutinominguez, wim.vanrossum\}@tno.nl}
}

%
%
\ifx\arXivPrint\undefined\else

\makeatletter
\twocolumn[{
\vspace{2cm}
This paper has been accepted for publication at the

\vspace{1cm}
\centerline{\textbf{\huge{2023 IEEE Radar Conference}}}

\vspace{5cm}

\vspace{1cm}
\textbf{Citation}\\
P.B. Cox, M.A. Coutino, and W.L. van Rossum, ``\doSingleLine{\@title},'' in \textit{Proceedings of the 2023 IEEE Radar Conference}, pp --, San Antonio, Texas, USA, May 2023.

\vspace{1cm}

\definecolor{commentcolor}{gray}{0.9}
\newcommand{\commentbox}[1] {\colorbox{commentcolor}{\parbox{\linewidth}{#1}}}

\vspace{4cm}
\commentbox{
	\vspace*{0.2cm}
	\hspace*{0.2cm}More papers from P.B. Cox can be found at\\~\\
	\centerline{\large{\url{https://orcid.org/0000-0002-8220-7050}}}\\~\\
	\hspace*{0.2cm}and of M.A. Coutino at\\
	\centerline{\large{\url{https://scholar.google.com/citations?user=APLpE9cAAAAJ}}}\\~\\
	\hspace*{0.2cm}and of W.L. van Rossum at\\
	\centerline{\large{\url{https://scholar.google.com/citations?user=Lh1u0qMAAAAJ}}}\\~\\
	\vspace*{0.2cm}
}

\vspace{3cm}
\textcopyright 2023 IEEE. Personal use of this material is permitted. Permission from IEEE must be obtained for all other uses, in any current or future media, including reprinting/republishing this material for advertising or promotional purposes, creating new collective works, for resale or redistribution to servers or lists, or reuse of any copyrighted component of this work in other works.
}]
\clearpage
\makeatother

\fi
%
%

\maketitle

\begin{abstract}
Efficient clutter filtering for pulsed radar systems remains an open issue when employing pulse-to-pulse modulation and irregular pulse interval waveforms within the coherent processing interval. The range and Doppler domain should be jointly processed for effective filtering leading to a large computational overhead. In this paper, the joint domain filtering is performed by constructing a clutter projection matrix, also known as the projected \emph{non-identical multiple pulse compression} (NIMPC) method. The paper extends the projected NIMPC filter to irregular pulse interval waveforms. Additionally, a kernel-based regularization will be introduced to tackle the ill-conditioning of the matrix inverse of the NIMPC method. The regularization is based on a model of the second-order statistics of the clutter. Moreover, a computationally efficient algorithm is formulated based on fast Fourier transforms and the projected conjugate gradient method. Through a Monte Carlo study it is demonstrated that the proposed kernelized filtering outperforms the projected NIMPC in clutter filtering.
\end{abstract}


\section{Introduction}

Waveform agility combined with modern digital signal processing has the potential to significantly increase the flexibility of modern radar systems~\cite{Gurbuz2019}. These agile waveforms allow for more design freedom to adapt to specific tasks or environmental conditions when combined with appropriate processing techniques~\cite{Wicks2011,Aubry2016}. When range and Doppler are processed in a decoupled way, the irregular pulse modulation and/or irregular pulse interval leads to \emph{range sidelobe modulation} (RSM)~\cite{Blunt2010,Sahin2017} that can significantly impact the clutter cancellation filters.

A potential approach to reduce RSM for clutter filters is the design of \emph{mismatched filters} (MMFs)~\cite{Ackroyd1973,Stoica2008}. The MMFs design for irregular waveforms is a trade-off between a lower RSM at the cost of higher overall range sidelobes, which can become conservative for pulse diverse waveforms~\cite{Blunt2020}.

Alternatively, irregular waveforms can be jointly processed in the range-Doppler domain for the \emph{coherent processing interval} (CPI) using, e.g., sparse signal processing, \emph{projected non-identical multiple pulse compression} (NIMPC), iterative filters~\cite{Yardibi2010,Rossum2020,DeMartin2020,Jones2021}. Unfortunately, these methods can have a tremendous computational overhead and the filtering problem can be ill-conditioned. 

In this paper, a joint domain clutter filter for irregular waveforms is presented that tackles the ill-conditioning and significantly alleviates the computational overhead. More specifically, a kernel-based regularization will be introduced to improve conditioning of the large scale matrix inverse, similar to~\cite{Chen2012,Darwish2018}. The regularization is based on second-order statistics of the clutter using clutter models~\cite{Aubry2012,Aubry2013,Ward2013,Wu2020}. During operation, the statistics of the clutter can be predicted by utilizing digital terrain maps and RCS clutter models.

To alleviate the computational burden, in this paper, a computationally efficient algorithm is formulated based on \emph{fast Fourier transforms} (FFTs) and the \emph{projected conjugate gradient} (PCG) method, similar to \cite{Jones2021}. The relation of our method with respect to the \emph{extensive cancellation algorithm} (ECA)~\cite{Colone2009} and NIMPC~\cite{Jones2021} will be discussed.

Summarizing, the paper will present and analyze a computationally attractive clutter filtering technique for irregular waveforms. The contributions of this paper are:
\begin{enumerate*}[label=(\roman*)]
	\item a kernelized regularization framework for clutter filtering is introduced,
	\item a design methodology is discussed to formulate the kernel (regularizer) by modeling the second-order statistics of the clutter,
	\item a computationally attractive algorithm using FFTs and PCG is derived, and
	\item a clutter filtering strategy for waveforms with irregular pulse intervals and pulse-to-pulse modulation is defined.
\end{enumerate*}

The paper is organized as follows. In Sec.~\ref{sec:signal model}, the signal model is introduced. The clutter filtering with target estimation problem is defined in Sec.~\ref{sec: clutter rejection}. The design of the kernel for the clutter is discussed in Sec.~\ref{sec:correlation clutter}. The computational efficient implementation is described in Sec.~\ref{sec:decreasing computational complexity}. In Sec.~\ref{sec:examples}, the effectiveness of the clutter filter is demonstrated by examples, followed by the conclusions in Sec.~\ref{sec:conclusion}.

\section{Signal model} \label{sec:signal model}

In the paper, the received signal at baseband $\by\in\Cmplx^D$ is assumed to contain a target signal component $\bs\in\Cmplx^D$, clutter signal component $\bc\in\Cmplx^D$, and a circular Gaussian noise $\be\sim\CNormal(0,\sigma^2 \bm I)$ combined as $\by=\bs+\bc+\be$. The target and signal components are modeled as~\cite{Rosenberg2022}
\begin{equation}
\by = \bAxt \bxt + \bAxc\bxc + \be,
\label{eq:linear system model}
\end{equation}
where $\bAxt\in\Cmplx^{D\times F}$ and $\bAxc\in\Cmplx^{D\times G}$ are the linear models of the target(s) and clutter, respectively, and $\bxt\in\Cmplx^F$ and $\bxc\in\Cmplx^G$ denote the returns of the target and clutter, respectively. The matrices $\bAxt, \bAxc$ can represent different range-Doppler domains and their columns are composed of time-shifted (range), time-dilated, and Doppler-shifted versions of the transmitted waveform $s(t)$~\cite{Kelly1965}\footnote{The constant phase shift $\exp(j2\pi f_c\tau_j)$ induced by the Doppler effect is not included in the linear model~\eqref{eq:Columns of linear signal model} as it can be absorbed in $\bxt$ or $\bxc$.}, i.e.,
\begin{equation}
[\bA]_{i,j\vert(\tau_j,v_j)} = s(\alpha_j(k_i-\tau_j))\exp(j 2\pi f_c (1-\alpha_j)k_i),
\label{eq:Columns of linear signal model}
\end{equation}
where $[\cdot]_{i,j}$ defines the $(i,j)$-th element of the matrix, $k_i$ is the $i$-th element in the discrete time vector $\bk=[0,…,(N-1) f_s]$ with sampling frequency $f_s\in\Real^+$, the columns of $\bAxt$, $\bAxc$ represent some $(\tau_j,v_j)$ pair related to the range $R_j=\frac{1}{2}\tau_jc$ and velocity $v_j\in\Real$ of a target or clutter response, $f_c\in\Real$ is the carrier frequency, $\alpha_j\approx 1-2\frac{v_j}{c}$ denotes the Doppler stretch factor, and $c\in\Real$ is the speed of light in vacuum. 

The matrices $\bAxt$ and $\bAxc$ may represent different range-Doppler domains where clutter or target responses are expected, e.g., the clutter velocity is expected up to $\vert v\vert\leq30$\,m/s and the target velocity up to $\vert v\vert\leq400$\,m/s. If no such differentiation can be made, then $\bAxt=\bAxc=\bA$.

In the linear model~\eqref{eq:linear system model}-\eqref{eq:Columns of linear signal model}, the following considerations are been made:
\begin{enumerate*}[label=(\roman*)]
	\item  $v_i\ll c$ $\forall v_i$ (approximation for the Doppler stretch factor $\alpha$)~\cite{Kelly1965},
	\item the target and clutter have constant velocity during CPI, and 
	\item the clutter returns are coherent during the listening time.
\end{enumerate*}

Alternatively, the clutter matrix $\bAxc$ has also been represented by other (orthonormal) bases, such as tuned Q wavelet transform or short time Fourier transform, e.g., see~\cite{Rosenberg2022}. In this paper, the model~\eqref{eq:Columns of linear signal model} is used as well-known clutter models can be used to design the kernel, see Sec.~\ref{sec:correlation clutter}.

\section{Clutter rejection and target estimation} \label{sec: clutter rejection}

Based on the defined signal model~\eqref{eq:linear system model}, the clutter cancellation and target estimation problem will be formulated in this section. Generally, the $(\tau_j,v_j)$ pairs are selected based on a fixed linear grid given by $\bm\tau=\bk$ and $\bm v=\left[v_{min}~~v_{min}+\Delta v~~\ldots~~v_{max}\right]^\top$. The chosen grid points should coincide with the Doppler resolution of the transmitted waveform where, as a rule of thumb, $\Delta v= \frac{c}{2T_{obs}f_c}$ with observation time $T_{obs}$. The number of $(\tau_j,v_j)$ pairs and, therefore, the number of elements in $\bxt$ and $\bxc$ can be large. Generally speaking, the number of non-zero elements in $\bxt$ is limited, leading to a sparse reconstruction problem w.r.t. $\bAxt\bxt$. On the other hand, the clutter surfaces or volume clutter can have high returns in an extended range-Doppler region. Hence, in this paper, the joint clutter rejection and target estimation problem is solved by
\begin{equation}
\min_{\bxt,\bxc} \!\left\Vert\by\!-\![\bAxt~\bAxc]\!\left[\begin{array}{c}\bxt\\\bxc\end{array}\right]\right\Vert_2^2 \!+\! \lambda_T\Vert\bxt\Vert_1^2\!+\! \lambda_C\Vert\bxc\Vert_{\bSigmac}^2,
\label{eq:mixed l2 l1 problem}
\end{equation}
where $\Vert\cdot\Vert_k$ is the $\ell_k$-induced signal norm, $\Vert\cdot\Vert_{\bSigma}^2$ is the squared weighted $\ell_2$-norm, i.e., $\Vert x\Vert_{\bSigma}^2=\bx^H\bSigma^{-1}\bx$ with positive definite (symmetric) weighting $\bSigma$. In general, the problem~\eqref{eq:mixed l2 l1 problem} without regularization, i.e., $\lambda_T=\lambda_C=0$, will be ill-posed as, for radar systems, the number of unknowns is larger than the number of samples leading to many possible solutions. In~\cite{Chen2012,Darwish2018}, it has been demonstrated that adding weighted regularization term, the estimator can decrease its overall \emph{mean squared estimation error} (MSE) by trading a small amount of the estimation bias to largely decrease the variance of the estimator, which is known as the bias-variance trade-off.

One could argue to solve the ill-conditioning issue of~\eqref{eq:mixed l2 l1 problem} by simply increasing the observation time $T_{obs}$, e.g., by increasing the dwell with more pulses. However, the necessary velocity grid distance $\Delta v$ for $\bAxt$ and $\bAxc$ is inversely proportional to $T_{obs}$, i.e., $\Delta v \sim 1/T_{obs}$. In other words, an increased observation time requires a finer velocity grid $\bm v$ and, hence, an increased number of the to-be-estimated parameters.

In case that the influence of the target responses on the estimation of the clutter are negligible, e.g., when detecting weak targets in the presence of strong clutter, then the problem in~\eqref{eq:mixed l2 l1 problem} can be solved in subsequent steps. First, the clutter filtering step is performed
\begin{equation}
\by_{filt} = \underbrace{\left(I-\bAxc\left(\bAxc^H\bAxc+\lambda_C\bSigmac^{-1}\right)^{-1}\bAxc^H\right)}_{\bP(\lambda_C\bSigmac)}\by,
\label{eq:clutterFiltering}
\end{equation}
followed by sparse target reconstruction step
\begin{equation}
\min_{\bxt} \left\Vert\by_{filt}-\bAxt\bxt\right\Vert_2^2 + \lambda_T\Vert\bxt\Vert_1^2.
\label{eq:target estimation}
\end{equation}

In case that the target responses on the estimation of the clutter are not neglectable, then the solution to~\eqref{eq:mixed l2 l1 problem} could be found by iterating between~\eqref{eq:clutterFiltering} and~\eqref{eq:target estimation}. At the $k$-th iteration, $\by$ in~\eqref{eq:clutterFiltering} should be replaced by $\by-\bAxt\bx_{T,k}$ leading to $\by_{filt,k}$. Then, $\bx_{T,k+1}$ is obtained by solving~\eqref{eq:target estimation} based on $\by_{filt,k}$ instead of $\by_{filt}$. As the problem is jointly convex, alternating minimization will converge to the optimal solution.

In~\cite{Colone2009,Jones2021}, the orthogonal projector $\spanS(\bAxc)^\perp$ has been used to formulate the ECA and Proj-NIMPC methodologies which is equivalent to setting $\lambda_C=0$ ($\bP(0)$) in~\eqref{eq:clutterFiltering}.

\section{Kernel of the clutter} \label{sec:correlation clutter}

In this section, a formulation for the correlation of the clutter will be discussed based on~\cite{Aubry2012,Aubry2013,Wu2020}. Note that there exists a vast literature on modeling clutter which will not be treated here. During operation, the correlation can be predicted by utilizing digital terrain maps and RCS clutter models or clutter map estimation techniques, see~\cite{Aubry2012,Aubry2013} for a detailed discussion. The presented clutter filtering approach could be applied for various formulations of clutter or interference. Therefore, the kernel formulation is a general framework to treat clutter or interference filtering for irregular waveforms.

Similar to~\cite{Aubry2012,Aubry2013,Ward2013,Wu2020}, the mean amplitude of the clutter scatterers is\footnote{If it is assumed that the phase of the clutter returns are uniformly distributed on $[-\pi,\pi]$ then $\Expct[\bxc] = 0$. Moreover, a non-zero mean can be assumed in the estimation process, however, adequately parameterizing the covariance matrix should be sufficient to avoid parameterizing the mean~\cite{Rasmussen2006}.}
\begin{equation}
\Expct[\bxc] = \bm 0,
\label{eq:clutter zero mean}
\end{equation}
and the covariance matrix is parameterized as
\begin{equation}
\bSigmac = \Expct[\bxc\bxc^H] \!=\! \diag\left(\!\big[\begin{array}{ccc}\sigma^2_{(\tau_1,v_1)}\!&\!\ldots\!&\!\sigma^2_{(\tau_M,v_M)}\end{array}\big]\!\right),
\label{eq:clutter covariance}
\end{equation}
where
\begin{subequations} \label{eq: covariance}
\begin{align}
\sigma^2_{(\tau_j,v_j)} &= \sigma^{(\tau_j)}_0\frac{P_tG\lambda^2}{(4\pi)^3\left(\frac{1}{2}c\tau_j\right)^4L_s}  \times \label{eq:covariance range} \\
 &\qquad\qquad\int_{v_j-\frac{\Delta v}{2}}^{v_j+\frac{\Delta v}{2}} \exp\left[ -\frac{(s-v_c)^2}{2\sigma_s^2}\right] \rm{d}s,  \label{eq:covariance Doppler}
\end{align}
\end{subequations}
where $\sigma^{(\tau_j)}_0\in\mathbb{R}^+$ is the RCS of the clutter at range bin $\tau_j$, $P_t\in\mathbb{R}^+$ is the transmit power, $G\in\mathbb{R}^+$ is the antenna gain, $\lambda\in\mathbb{R}^+$ defines the wavelength of the carrier frequency, $L_s\in\mathbb{R}^+$ are all system losses, and $v_c$ is the average radial speed of the clutter. The radar range equation is visible in~\eqref{eq:covariance range} to determine the power over the Doppler bins $v_j$ and~\eqref{eq:covariance Doppler} defines a Gaussian Doppler spectrum for range bin $\tau_j$.

Note that $\bSigmac$ represents a spatial correlation and it does not represent a time correlation. In this paper, we assume that the various $(\tau_j,v_j)$ pairs are spatially uncorrelated , i.e., $\bSigmac$ in \eqref{eq:clutter covariance} has no off-diagonal terms. If desired, the off-diagonal terms can be included without loss of generality.

In this paper, we focus on sea surface clutter returns for which the RCS $\sigma^{(\tau_j)}_0$ at the $\tau_j$-th range bin is modeled by
\begin{equation}
\sigma^{(\tau_j)}_0 = \frac{10^{0.6K_b\sin\psi^{(\tau_j)}}}{2.51\cdot10^6\lambda} A^{(\tau_j)}, \label{eq:sea sureface}
\end{equation}
where $A^{(\tau_j)}$ is the area of the $\tau_j$-th range bin, $\psi^{(\tau_j)}$ denotes the grazing angle, and $K_b$ is the constant on the Beaufort wind scale. Models for $\sigma^{(\tau_j)}_0$ exist for other clutter types, e.g., hilly ground clutter or rain/snow volume clutter~\cite{Aubry2012,Aubry2013,Wu2020}.

Combining~\eqref{eq: covariance} and~\eqref{eq:sea sureface} will lead to the covariance matrix $\bSigmac$~\eqref{eq:clutter covariance}. We would like to stress that any other model or estimate of the clutter covariance can be used without loss of generality. When the clutter $\bxc$ is defined by~\eqref{eq:clutter zero mean}-\eqref{eq:clutter covariance}, then the minimal MSE estimator of $\bxc$ in~\eqref{eq:mixed l2 l1 problem} is obtain when using the covariance as the symmetric weighing~\cite{Chen2012,Darwish2018}.

\section{Decreasing the computational complexity}\label{sec:decreasing computational complexity}

In this section, a computationally efficient algorithm is presented inspired by~\cite{Colone2009,Higgins2011,Jones2021}. For the clutter filter, the following assumptions are taken
\begin{enumerate}[label=A\arabic*]
	\item the pulse time stretching caused by the Doppler stretch factor is negligible.\label{A:time stretch pulse} 
	\item the intrapulse Doppler is negligible for the clutter. \label{A:intrapulse Doppler}
	\item the largest range bin to filter the clutter is smaller than unambiguous range of each transmitted pulse. \label{A:pulse decomposistion}
	\item the pairs $(\tau_j,v_j)$ represent a uniform rectangular grid in $\tau$ and $v$ given by $(\tau_j,v_k)$ with $j=1,\ldots J,k=1,\ldots,K$. \label{A:Uniform grid}
\end{enumerate}

The assumptions~\ref{A:time stretch pulse} and~\ref{A:intrapulse Doppler} are well-known assumptions, which are valid when the clutter has a relatively low velocity and the transmitted waveform is narrow-band. Assumption~\ref{A:pulse decomposistion} allows to apply clutter filtering on the individual pulses only and, therefore, the FFT is computed of the single pulses instead of the CPI. \ref{A:pulse decomposistion} is valid when clutter is unambiguous in range on all pulses. Assumption~\ref{A:Uniform grid} is taken to simplify the notation.

Under~\ref{A:time stretch pulse}-\ref{A:Uniform grid}, the clutter contributions and noise contributions on the  $L\times1$ received vector for the $m$-th pulse are
\begin{equation} \label{eq:Simplified model}
   \by_{C,m} = \underbrace{\left[\bphi_m^\top \otimes \bS_m \right]}_{\bA_{C,m}} \bx_C+\be_m,
\end{equation}
where all Doppler phases are collected as $\bphi_m =[ \begin{array}{ccc} e^{j2\pi T_{s,m} \frac{2v_1}{c}f_c} & \hdots & e^{j2\pi T_{s,m} \frac{2v_K}{c}f_c}\end{array} ]^\top$, $T_{s,m}$ is the time of starting to transmit pulse $m$ and $T_{s,1}=0$, $\bx_{C}\in\Cmplx^{K(L-N+1)}$ is the to-be-estimated clutter responses, $\be_m\in\Cmplx^{(L-N+1)}$ is a realization of a white circularly symmetric Gaussian noise, $\otimes$ denotes the Kronecker product, $L\leq\min(f_sT_{d,m})$ $\forall m$ is an integer, $T_{d,m}$ is the pulse interval between the $m$-th and $m+1$-th pulse\footnote{\label{footnote:sampling grid}It is assumed that $T_{d,m}$ and $\tau_{p,m}$ are on the sampling grid.}, and $\bS_m\in\Cmplx^{L\times(L-N+1)}$ denotes the Toeplitz matrix where the first column is given by
\begin{equation}
     \bs_m = \left[\begin{array}{cccccc} s_{m,1} & \hdots & s_{m,N} & 0 & \hdots & 0 \end{array} \right]^\top,
\end{equation}
where $s_{m,n}$ is the $n$-th discrete sample of the $m$-th waveform $s_m(t)$ with $N=\max(f_s\tau_{p,m})$ as an integer and $\tau_{p,m}$ is the pulse length of the $m$-th pulse\cref{footnote:sampling grid}. The clutter and noise model~\eqref{eq:Simplified model} connect back to~\eqref{eq:linear system model} by concatenating the $m$-th pulses column-wise and including appropriate zero-padding between the pulses. The samples of $\by$ that will be clutter filtered are concatenated in $\tilde{\by}$.

Next, the structure in $\bA_{C,m}$ is used to simplify the computation of~\eqref{eq:clutterFiltering}. Starting from the right in~\eqref{eq:clutterFiltering}, see that,
\begin{multline}
\bf = \bA_C^H\tilde{\by} = \sum_{m=1}^M \left[ \bphi^*_m\otimes \bS_m^H \right]\tilde{\by}_m \\
	= \sum_{m=1}^M \vectorize\left\{ \bF^H\left[ (\bF \bs_m)^* \odot \bF\tilde{\by}_m \right] \bphi^H_m \right\},
\label{eq:Left Matrix Vector product}
\end{multline}
where $\vectorize\{\bA\}$ defines vectorization of a matrix $\bA$, $\bF$ is the discrete Fourier transform matrix, $\bx^*$ denotes the complex conjugate of $\bx$, and $\odot$ is the Hadamard product. Note that in~\eqref{eq:Left Matrix Vector product}, the Toeplitz structure is used to simplify the matrix-vector product by element-wise vector operations of the FFT of the vectors. Then, moving further through~\eqref{eq:clutterFiltering}, define $\bg$ as the solution to
\begin{equation}
\bg = \left(\bAxc^H\bAxc+\lambda_C\bSigmac^{-1}\right)^{-1} \bf.
\label{eq:inverse within Filter}
\end{equation}
The solution to the inverse can, for example, be found by solving the PCG method~\cite{Nocedal2006}. The PCG obtains $\bg$ in~\eqref{eq:inverse within Filter} iteratively without directly computing the inverse. At each iteration of the PCG, the product $\left(\bAxc^H\bAxc\!+\!\lambda_C\bSigmac^{-1}\right) \bg_q$ is computed where $\bg_q$ is the solution of $\bg$ in~\eqref{eq:inverse within Filter} at iteration $q$. Advantageously, this product can efficiently be computed using matrix-vector multiplications involving FFTs as
\begin{allowdisplaybreaks}
\begin{multline}
    \left(\bAxc^H\bAxc\!+\!\lambda_C\bSigmac^{-1}\right) \bg_q =\!  \sum_{m=1}^M\!\left(\bphi^*_m\bphi^\top_m \!\otimes\! \bS^H_m\bS_m \right) \bg_q + \bsigma\odot\bg_q \\
    = \sum_{m=1}^M \vectorize\left\{ \bS^H_m\bS_m \left[\begin{array}{ccc} \bg_{q,1} \!&\! \hdots \!&\! \bg_{q,K}   \end{array}\right]\bphi_m^H\bphi_m \right\}+ \bsigma\odot\bg_q \\
		=  \bsigma\odot\bg_q + \sum_{m=1}^M\sum_{k=1}^K \vectorize\Big\{ \bF^H\Big( (\bF \bs_m)^*\odot\bF s_m\\ \odot\bF\bg_{q,k} \Big)\left[\bphi_m\right]_k\bphi_m^H\Big\},
		\label{eq:PCG iter matrix-vector}
\end{multline}
\end{allowdisplaybreaks}
where $\bsigma=\lambda_C[\begin{array}{ccc}\sigma^{-2}_{(\tau_1,v_1)}\!&\!\ldots\!&\!\sigma^{-2}_{(\tau_J,v_K)}\end{array}]^\top$ and $\bg_{q,k}\in\Cmplx^{L-N+1\times 1}$ is the $k$-th block in $\bg_q$. After convergence of the PCG and obtaining $\bg$, the filtered output can be found by
\begin{subequations}\label{eq:clutter prediction}
\begin{multline}
\tilde{\by}_{filt}=\tilde{\by}-\bAxc\bg = \tilde{\by}- \vectorize\big\{  \big[ \bS_1\left[\begin{array}{ccc}\bg_1&\! \ldots\!&\bg_K\end{array}\right]\!\bphi_1 
\\\begin{array}{cc}\hdots&\bS_M\left[\begin{array}{ccc}\bg_1&\!\ldots\!&\bg_K\end{array}\right]\bphi_M \end{array}\! \big] \big\},
\label{eq:clutter prediction vector}
\end{multline}
where
\begin{multline}
\bS_m\left[\begin{array}{ccc}\bg_1&\ldots&\bg_K\end{array}\right]\bphi_m = \\
\sum_{k=1}^K \bF^H \left(\bF s_m\odot \bF\tilde{\bg}_k\right)\left[\bphi_m\right]_k.
\label{eq:clutter forward matrix}
\end{multline}
\end{subequations}

The regularization term in~\eqref{eq:PCG iter matrix-vector} adds $2K(L-N+1)$ operations per iteration to the PCG. The total number of operations to compute~\eqref{eq:clutterFiltering} using the PCG is $4(2I+1)KML\log_2(L)+2IK(L-N+1)$ with $I$ the total number of iterations of the PCG, see, e.g.,~\cite{Jones2021} for more details on the computational complexity of proj-NIMPC. If the inverse is directly computed in~\eqref{eq:inverse within Filter} instead of the PCG, then the number of operations is $K^3(L-N+1)^2 + 2K^2M(L-N+1)L\log_2(L)+K(L-N+1)$. Hence, using~\eqref{eq:Left Matrix Vector product}, the PCG with matrix-vector multiplication~\eqref{eq:PCG iter matrix-vector}, and~\eqref{eq:clutter prediction}, the clutter filtering problem in~\eqref{eq:clutterFiltering} can be solved in a computationally efficient manner.

\section{Simulation examples} \label{sec:examples}

In this section, the effectiveness of the proposed clutter filtering strategy is demonstrated. The applied waveform is composed of identical \emph{linear frequency modulated} (LFM) pulses with bandwidth $B=5$\,MHz and a pulse duration of $\tau_p=40$\,\textmu s transmitted at an irregular interval drawn from a uniform distribution $\mathcal{U}\left(500~800\right)$\,\textmu s rounded on the sampling grid. The sampling frequency is $f_s=10$\,MHz and the center frequency is $f_c=10$\,GHz. The projected conjugate gradient \texttt{pcg} routine of Matlab 2020b is used with an absolute tolerance of $10^{-13}$. The diagonal block of the block-circulant matrix $\bAxc^H\bAxc$ is applied as a preconditioner of the \texttt{pcg}. In the following examples, it is assumed that $\bSigmac$ is known. In real applications, the unknown parameters defining the kernel $\bSigmac$ may be estimated from data using marginal likelihood optimization~\cite{Chen2012,Darwish2018,Rasmussen2006}\footnote{Other covariance estimation techniques may be used, e.g.,~\cite{Steiner2000,DeMaio2019}. However, these techniques will not utilize the kernel model defined in Sec.~\ref{sec:correlation clutter}.}.

\subsection{Designing the clutter filter}
\begin{figure}[!t]%
	\centering
	\input{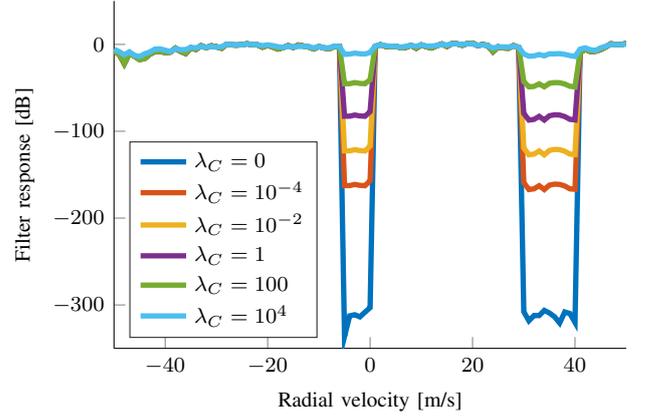}
	\caption{The filter response to a unit input for a clutter filter designed with $v_k=\left\{-5, -4, \ldots, 0\right\}\cup\left\{30, 31, \ldots, 40\right\}$\,m/s, covariance matrix $\bSigmac=I$, and range bin $6.43$\,km is selected. The waveform contains $N_p=32$ pulses.}%
	\label{fig:influence of regularization}%
	\vspace{-0.5cm}
\end{figure}

First, the influence of the regularization term $\lambda_C$ will be demonstrated. The clutter filter is designed with a velocity grid $v_k=\left\{-5, -4, \ldots, 0\right\}\cup\left\{30, 31, \ldots, 40\right\}$\,m/s, the clutter covariance matrix is $\bSigmac=I$, and the waveform is composed of $N_p=32$ LFM pulses with irregular PRI. Fig.~\ref{fig:influence of regularization} shows the filter response at a range bin $6.43$\,km. The response at other range bins is similar. The figure shows dips at the expected locations corresponding to $v_k$. For larger values of $\lambda_C$, the regularization $\lambda_CI$ term will dominate compared to $\bAxc^H\bAxc$ in~\eqref{eq:PCG iter matrix-vector} and, hence, the influence of the designed clutter filter in $\bAxc^H\bAxc$ decreases as expected.

\begin{figure}[!t]%
	\centering
	\input{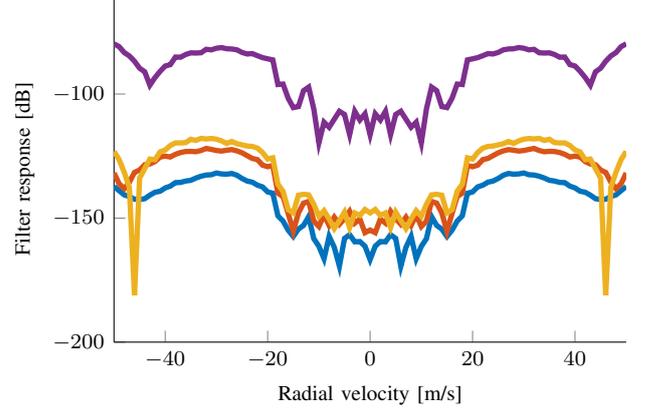}
	\caption{The filter response to a unit input for various realizations of the irregular PRI waveform with $N_p=32$ pulses is shown where the clutter filter is designed with a grid $v_k=\left\{-18, -17, \ldots, 18\right\}$\,m/s, regularization $\lambda_C\bSigmac=10^{-4}I$, and range bin $6.43$\,km is selected.}%
	\label{fig:realization of the filter}%
	\vspace{-0.5cm}
\end{figure}

The proposed clutter filter strategy will not have sharp edges as shown in Fig.~\ref{fig:influence of regularization} in all cases. Fig.~\ref{fig:realization of the filter} shows the clutter filter response of a filter with grid $v_k=\left\{-18, -17, \ldots, 18\right\}$\,m/s and regularization $\lambda_C\bSigmac=10^{-4}I$ for various realizations of the irregular PRI waveform with $N_p=32$. Fig.~\ref{fig:realization of the filter} shows that the $\spanS(\bP(\lambda_C\bSigmac))$ and $\spanS(\bAxt)$ may not be orthogonal by design and, hence, the filter suppress objects in other velocity regions. In Fig.~\ref{fig:realization of the filter}, the suppression is -150\,dB to -90\,dB in regions outside $v_k$. The impact can be minimized by minimizing their common span, e.g., by $\min\Vert\bAxt^H\bP(\lambda_C\bSigmac)\Vert_F$, by appropriately choosing the pairs $(\tau_j,v_k)$ to construct $\bAxc$, by designing $\lambda_C\bSigmac$, and/or by designing the pulse interval and pulse modulation in the waveform.

\subsection{Reconstruction of the clutter} \label{subsec:clutter reconstruction}

In this example, the ability of the clutter filter to reconstruct the clutter signal and the effect of regularization on the reconstructability is assessed.  The sea clutter is simulated using radar constant $k_{radar} = \frac{P_tG\lambda^2}{(4\pi)^3L_s}=250\cdot 10^8\,\mbox{Wm}^2$ in~\eqref{eq:covariance range} with grazing angle $\psi=0.5\pi$, $K_b=5$, and beam width $\theta_{BW}=4^\circ$ in~\eqref{eq:sea sureface}. The average radial speed of the clutter is $v_c=-2.2$\,m/s and the variance is $\sigma_s^2=5$\,m/s in~\eqref{eq:covariance Doppler}. The performance is based on a Monte Carlo simulation study with $N_{MC}=100$ runs. At each run, a new realization of the additive white noise $\be$ in~\eqref{eq:linear system model} is drawn and the variance is selected to obtain a signal-to-noise ratio of $20$\,dB. Similarly, a new realization of the irregular pulse interval is drawn at each new run. The reconstructability is measured in terms of the best-fit-rate (BFR), i.e.,
\begin{equation}
\BFR=\max\left\{ 1-\frac{\sum_{k=1}^N\left\Vert [\bc]_k-[\hat{\by}_C]_k\right\Vert_2}{\sum_{k=1}^N\left\Vert[\bc]_k-\bar{\bc}\right\Vert_2} ,0 \right\}\cdot100\%,
\label{eq:}
\end{equation}
where $\bc$ is the noiseless clutter signal (see~\eqref{eq:linear system model}), $\hat{\by}_C=(I-\bP(\lambda_C\bSigmac))\by$ is the reconstruction of the clutter signal without noise\footnote{Note that $\hat{y}_C\!=\!(I\!-\!\bP\!(\lambda_C\bSigmac\!)\!)\by \!=\! \bAxc\!\left(\bAxc^H\bAxc\!+\!\lambda_C\bSigmac\right)^{-1}\!\bAxc^H\by$.}, and $\bar{\bc}$ defines the mean of $\bc$.

The simulation study is performed for varying number of pulses in the waveform $N_p=\{4,8,16,32,64,128\}$. The clutter is generated by point scatterers with covariance~\eqref{eq:clutter covariance} based on a uniform grid between $-30$\,m/s and $30$\,m/s. The grid spacing is selected as $\Delta v=\{7.5,4,2,1,0.5,0.25\}$\,m/s respectively to the number of pulses in the waveform. The grid $v_k$ of the clutter filter is matched to the grid of the clutter. The regularization parameter $\lambda_C$ is selected to obtain the highest BFR. For the first case with regularization, $\lambda_C=\{90,350,400,550,1000,2400\}$ is selected with $\Sigma_C=I$ and, for the second case, $\lambda_C=\{10,12,14,13,16,15\}$ is selected with $\Sigma_C$ based on the aforementioned clutter parameters.

\begin{figure}[!t]%
	\centering
	\includegraphics[width=0.43\textwidth, angle=0]{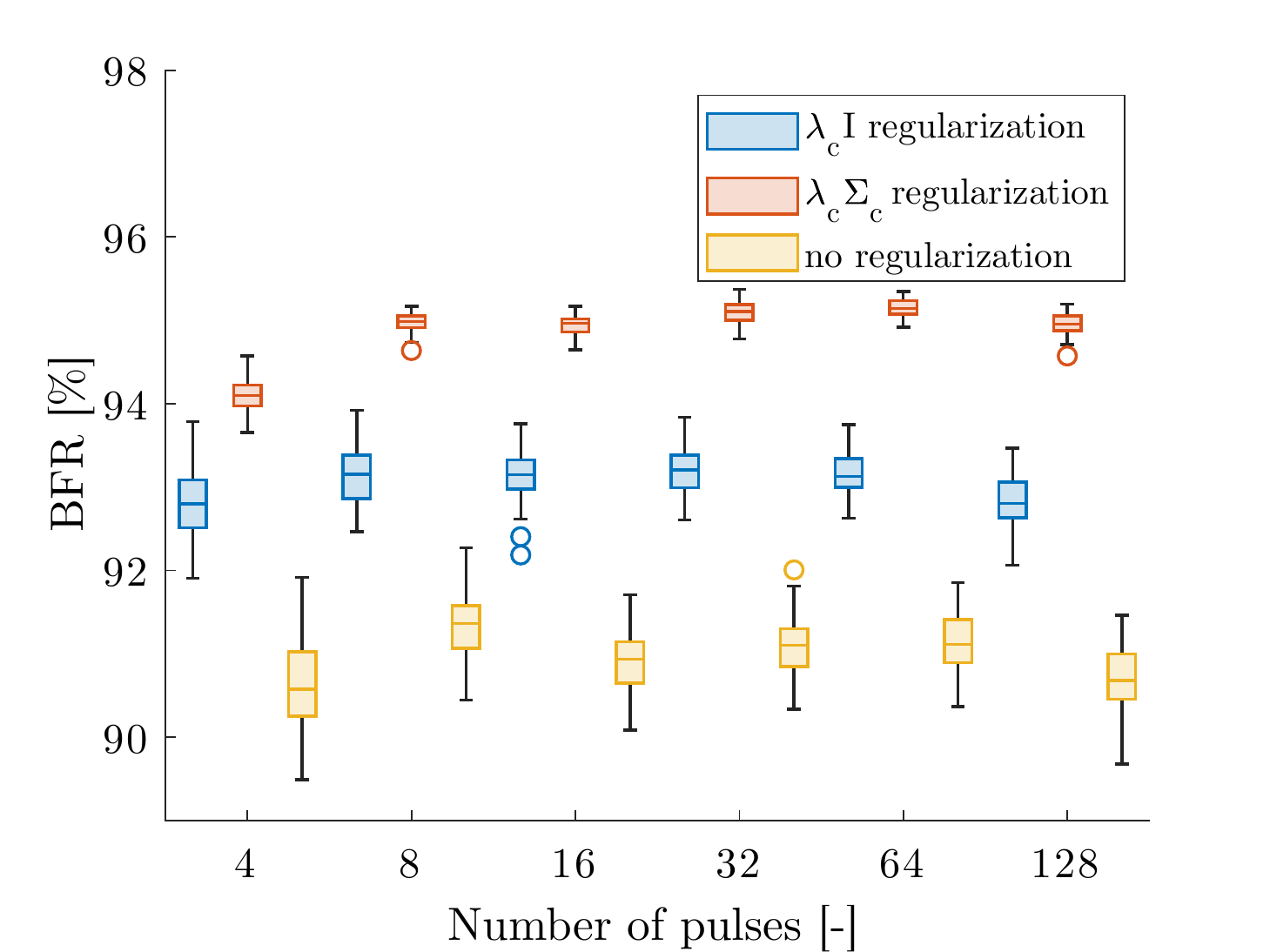}
	\caption{The BFR for various number of pulses in the waveform $N_p=\{4,8,16,32,64,128\}$ of the clutter filter without regularization, with $\lambda_CI$ regularization, and with $\lambda_C\bSigmac$ regularization for $N_{MC}=100$ Monte Carlo runs. The SNR is 20\,dB.}%
	\label{fig:BFR example}%
	\vspace{-0.5cm}
\end{figure}

\begin{figure}[!t]%
	\centering
	\input{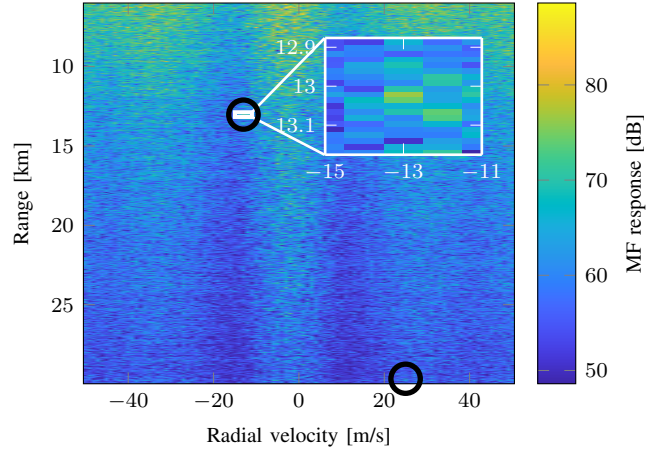}
	\caption{The matched filtered response of the signal with clutter and weak targets. The target locations are highlighted by black circles.}%
	\label{fig:clutter and target MF}%
	\vspace{-0.2cm}
\end{figure}

\begin{figure}[!t]%
	\centering
	\input{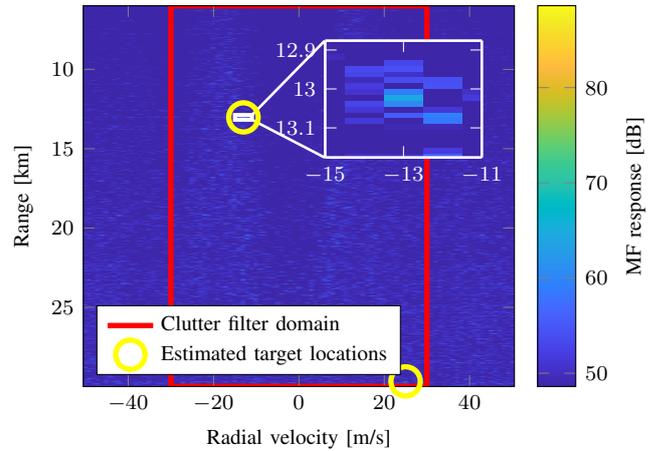}
	\caption{The matched filtered response of the signal after clutter filtering. The red frame indicates the domain where the parameters $\bxc$ of the clutter filter lie and the yellow circles highlight the estimated target locations.}%
	\label{fig:filtered MF}%
	\vspace{-0.5cm}
\end{figure}

Fig.~\ref{fig:BFR example} shows the BFR for various number of pulses for the case without regularization, with $\lambda_CI$ regularization, and with $\lambda_C\bSigmac$ regularization for $N_{MC}=100$ Monte Carlo runs. The figure highlights that adding a regularization significantly improves the estimation of the clutter as expected. The bias-variance trade-off is also visible from the figure, where the overall BFR\footnote{The BFR is related to the $\mbox{MSE}=\sum_{k=1}^N\left\Vert [\bc]_k-[\hat{\by}_C]_k\right\Vert_2$.} improves significant with a lower variance on the BFR when using regularization. Similarly, the regularization  $\lambda_C\bSigmac$ outperforms $\lambda_CI$ in terms of the BFR as $\bSigmac$ is the covariance of the to-be-estimated parameters which leads to the optimal bias-variance trade-off.

\subsection{Clutter filtering in the presence of weak targets}

The effectiveness of the clutter filtering in a scene with weak targets is shown next. The clutter coefficients $\bxc$ are sampled from an exponential distribution with average value given by the parameters highlighted in Sec.~\ref{subsec:clutter reconstruction} for a waveform with $N_p=32$ pulses. The scene will have two targets at $(13.03\,\mbox{km},-13\,\mbox{m/s})$ and $(30.71\,\mbox{km}, 25\,\mbox{m/s})$ with $|\bxt|=0.5$ and $|\bxt|=0.3$, respectively. The matched filtered response of the received signal is given in Fig.~\ref{fig:clutter and target MF}. The clutter power decreases with increasing range and the clutter is symmetric in velocity around the $-2.2$\,m/s. Also high sidelobes of the clutter are visible in the velocity domain for $\vert v\vert\geq20$\,m/s. The clutter clearly masks both targets.

The matched filtered response after clutter filtering is shown in Fig.~\ref{fig:filtered MF}. Additionally, Fig.~\ref{fig:filtered MF} highlights the estimated target locations by the $l_1$ orthonormal basis pursuit technique with two iterations, e.g., see~\cite{Struiksma2022}, using the signal after clutter filtering. The covariance matrix $\bSigmac$ lays more emphasis on filtering the clutter at close range and around a velocity of $-2$\,m/s, which can be observed by a dip in the matched filtered response. The sidelobes in the velocity domain of the clutter are now also removed which is most visible at close range. Also, the targets are no longer masked. Note that, the targets lie within the range-velocity domain of the clutter filter, see red frame in Fig.~\ref{fig:filtered MF}, however, the targets remain visible as the targets responses do not fit the clutter model in~\eqref{eq:clutter zero mean}-\eqref{eq:clutter covariance}. The estimated locations coincide with the actual locations and the peak power losses of the targets are 12.01\,dB and 5.736\,dB, respectively. To conclude, the clutter filtering strategy is capable of effectively removing the clutter for irregular waveforms and unmasking the weak targets.

\section{Conclusion} \label{sec:conclusion}

In this paper, a computationally efficient clutter cancellation filtering technique has been proposed for waveforms with irregular pulse intervals and pulse-to-pulse modulation within the coherent processing interval. More specifically, a kernel-based regularization has been introduced to elevate the ill-conditioning of the joint range-Doppler domain clutter estimation problem. The regularizer takes into account the second-order statistics of the clutter which prior knowledge can be based on digital terrain maps with clutter models and/or clutter map estimation techniques. The kernel-based regularization term steers the solution space of the clutter filter towards the hypothesized clutter statistics. Moreover, a computationally efficient methodology is formulated based on FFTs and the PCG method. The proposed clutter filtering strategy has been analyzed. The simulation study showed that adding a regularization term can significantly improve the clutter filtering process in terms of the best fit rate.

Obtaining an efficient technique that minimizes the impact of the clutter filtering in range-Doppler domain for target estimation remains a topic for future research. Also, a topic for future research is the automatized tuning of the covariance matrix based on data, which, e.g., could be achieved by marginal likelihood optimization.

\bibliographystyle{IEEEtran}

\bibliography{bibliography.bib}

\end{document}